\documentclass[%
 reprint,
 amsmath,amssymb,
 aps,
]{revtex4-1}

\usepackage{graphicx}
\usepackage{dcolumn}
\usepackage{bm}
\usepackage{braket}

\begin{document}


\title{Majorana zero energy modes in Silicene} 

\author{Anant V. Varma}
\affiliation{Indian Institute of Science Education and Research Kolkata, Mohanpur, 741246, West Bengal, India}

\author{Prasanta K. Panigrahi}
\affiliation{Indian Institute of Science Education and Research Kolkata, Mohanpur, 741246, West Bengal, India}

\date{\today}

\begin{abstract}
 Zero energy modes are shown to exist in silicene under suitably chosen magnetic as well as  electric fields. Two Majorana modes are found on application of opposite local magnetization on the silicene  sub-lattices. We identify a non-Majorana zero energy mode carrying pure spin current under the influence of inhomogeneous gate electric field. In both cases, wave functions reveal subtle interference pattern in phase space, showing structures finer than the Planck constant $\hbar$. A spin system coupled through spin-spin ($\sigma_{z}\otimes\sigma_{z}$) interaction with the Majorana modes exhibits periodic revival of coherence with a minimum period $\sim $1/n. Same system shows decoherence-free evolution in the case of gate electric field. Under a momentum dependent interaction one Majorana mode having bound state in the continuum character is found to be more robust as compared to the other. The mode arising on application of electric field shows rapid loss of coherence for such an interaction.

\end{abstract}

\maketitle


\section{Introduction}
 Realization of exotic field theoretical models in condensed matter systems (CMS) \cite{Sarma, shen, gang}, cold atoms \cite{Pachos, Rey, Ohberg} and optical set-ups \cite{Nori} have evoked strong interest in recent times. Appropriately engineered systems have led to experimental realization of relativistic field theoretical models in low energy regime \cite{Hughes}. Relativistic effects in (1+1) dimension like charge fractionalization in poly-acetylene \cite{Su,Jackiw,Triscone}, oxide hetero-structures \cite{Selvan}, chiral anomaly in charge-density-wave system \cite{Sakita}, optical realization of Jackiw-Rebbi model \cite{Priyam} and Majorana dynamics \cite{Rai} have been demonstrated. Planar systems like graphene \cite{Gordon} have exhibited relativistic quantum effects like Klein paradox \cite{Klein}, Zitterbewegung \cite{Zitter}, and realization of planar gauge \cite{Neto, Juan} and gravitation field \cite{Mohamad} theories. Interestingly, supersymmetric field theories have also manifested in CMS \cite{Hughes, Grover}, albeit awaiting experimental corroboration.  Another intriguing relativistic system, Majorana fermion in CMS has been in the forefront of current research \cite{Franz}. Initially evoked in modelling of neutrinos \cite{Ettore}, these exotic fermions, which are their own anti-particles have been proposed to exist in many combinations of systems e.g., semiconductor-superconductor \cite{Tewari}, topological insulator (TI)-superconductor  \cite{Sato} and  have been recently observed in insulator-topological superconductor system \cite{Zhang}. They have also been shown to exist in hexagonal systems as boundary states, in conjugation with superconductor \cite{EZA} and have found applications in error free quantum computation \cite{SDS,Hell,Vijay}. 

Here, we demonstrate existence of Majorana zero energy modes in the TI phase of silicene, under a local exchange field. Such a field on the two sublattices of silicene can be induced by sandwiching it between two ferromagnets with opposite magnetic ordering \cite{Moto, Long,Feng}. One of the Majorana mode is found to be more robust against loss of coherence and possesses bound state in the continuum (BIC) \cite{wigner} character. A non-Majorana zero energy state carrying a pure spin current is also  found for suitably chosen inhomogeneous gate electric field. Intriguingly, these modes are analogs of cat and kitten states in quantum optics and exhibit sub-Planck structure in phase space due to subtle quantum interference effects. When interacting with spins through spin-spin $\sigma_{z}\otimes\sigma_{z}$ interaction these Majorana modes exhibit periodic revival of coherence, owing to their entanglement properties.
While non-Majorana  mode provides decoherence free environment for a spin. Under momentum dependent interaction the Majorana mode with BIC character is more robust than the other. Whereas non-Majorana mode exhibits rapid loss of coherence. 

Silicene with a buckled hexagonal lattice structure has strong spin-orbit coupling (SOC): $H_{so} = \frac{\hbar}{4m_{0}^{2}c^{2}}( \overrightarrow{F} \times  \overrightarrow{p}).\overrightarrow{\sigma}$ leading to forces on the lattice plane. Forces parallel and perpendicular to the lattice plane, respectively induce intrinsic $(\lambda_{so})$ and Rashba $(\lambda_{r})$ SOC \cite{Ezawa}. The effective Hamiltonian in the low-energy regime, shows linear energy spectrum and is described by a next-nearest-neighbor tight-binding lattice model. Near the Dirac points it can be written as in the basis $(a_{\uparrow},b_{\uparrow},a _{\downarrow},b_{\downarrow})$: 

$\begin{bmatrix}
-\lambda_{so}+lE_{z} &\hbar v_{f}k_{+}  & ia\lambda_{r}k_{-} & 0\\ 
 \hbar v_{f}k_{-}& \lambda_{so}-lE_{z} &  0&-ia\lambda_{r}k_{-} \\ 
 -ia\lambda_{r}k_{-}& 0&\lambda_{so}+lE_{z}   &\hbar v_{f}k_{+} \\ 
0 & ia\lambda_{r}k_{+} & \hbar v_{f}k_{-} & -\lambda_{so}-lE_{z}
\end{bmatrix}$,

 with Fermi velocity $v_{f}= 5.5 \times  10^{5} m s^{-1}$, buckling separation of two sub-lattices along z-direction $l= 0.23 \AA $, intrinsic SOC parameter $\lambda_{so} = 3.9$ meV, lattice constant $a = 3.86 \AA $, Rashba SOC parameter $\lambda_{r}$ and $E_{z}(x,y)$ as gate electric field. We consider $\lambda_{r}=0$, as  band structure is not significantly influenced by $\lambda_{r}$ based on the tight-binding calculations \cite{ezaw}. For the present purpose we consider $E_{z}$ as function of only, and ansatz
$\Psi(x,y)= e^{ik_{y}y}\Phi(x)$, where $\Phi(x)$ is a four component spinor: $(a_{\uparrow},b_{\uparrow},a _{\downarrow},b_{\downarrow})$ is employed for . We also assume particle hole symmetry: $a_{\uparrow/\downarrow}= \pm ib_{\uparrow/\downarrow}$.

 In the following sections we identify the Majorana zero energy modes in the presence of magnetic field, and explicate their phase space structure. Section III is devoted to study of zero energy modes under gate electric field. Then interaction of n-qubit with these modes is investigated in section IV. After studying the loss of coherence of these modes under spin-spin and momentum dependent interaction we conclude with future direction of this research work.




\section{Majorana zero energy modes}

We now explicitly demonstrate the existence of Majorana zero energy modes in silicene. Majorana modes have been created here by including an $x$ dependent exchange energy term in the Hamiltonian $H_{J} = J(x) \hat{m}. \sigma \otimes \tau_{z}$ with $E_{z}=0$. The direction of magnetization is taken along z-axis:  $J(x) \sigma_{z}\otimes \tau_{z}$, where $\sigma_{z}$ acts on spin degree of freedom and $\tau_{z}$ acts on sub-lattice degree of freedom. Exchange energy on two sublattices can be induced by sandwiching silicene between two (different) ferromagnets. Inclusion of this term breaks time-reversal symmetry (TRS) in the system. For any arbitrary four component spinor two Majorana modes can be written as  \cite{Campos, Perti}:

\begin{center}
    $\Psi_{\pm} = \begin{pmatrix}
\psi_{1} \\ 
\psi_{2}\\ 
\psi_{3}\\ 
\psi_{4}
\end{pmatrix} \pm\begin{pmatrix}
-\psi_{4}^{*} \\ 
\psi_{3}^{*}\\ 
\psi_{2}^{*}\\ 
-\psi_{1}^{*}
\end{pmatrix} $\\
\end{center}

In our case, the zero energy solutions are found upon considering the components of spinor as: 

\begin{center}
    $\psi_{1} = e^{-(x-\alpha-i\beta)^{2}/\sigma^{2}} + e^{-(x+\alpha-i\beta)^{2}/\sigma^{2}} + \eta $ \\ 
    $\psi_{2}=i (e^{-(x-\alpha+i\beta)^{2}/\sigma^{2}} + e^{-(x+\alpha+i\beta)^{2}/\sigma^{2}} + \eta )$ \\ 
    
    $\psi_{3}= -i (e^{-(x-\alpha+i\beta)^{2}/\sigma^{2}} + e^{-(x+\alpha+i\beta)^{2}/\sigma^{2}})$ \\ 
    $\psi_{4}=-(e^{-(x-\alpha-i\beta)^{2}/\sigma^{2}} + e^{-(x+\alpha-i\beta)^{2}/\sigma^{2}})$ ,

\end{center}








where $\eta $ is a constant. One of the Majorana mode with positive sign can be explicitly written as $\Psi_{+} = \phi(x)(1, i, -i, -1)^{T}$, where $\phi(x)= e^{-(x-\alpha-i\beta)^{2}/\sigma^{2}} + e^{-(x+\alpha-i\beta)^{2}/\sigma^{2}} + e^{-(x-\alpha+i\beta)^{2}/\sigma^{2}} + e^{-(x+\alpha+i\beta)^{2}/\sigma^{2}}+ \eta  $. We notice that, for $\Psi_{+}$ the spatial part of the solution is separable, whereas for $\Psi_{-}$  it is not possible, unless $\eta$ is zero. There is a possibility of phase transition at this point for $\Psi_{+}$. State $\Psi_{+}$ is infact a bound state in continuum (BIC) \cite{David,Stone}, where $\eta ^{2}$ part forms a constant background. This state is normalizable in phase space, following the condition that conjugate momentum along x- direction $p_{x}\neq  0$. This state can be robust and decoherence free, as more recently predicted and experimentally realized in other systems \cite{Doug}. The other Majorana mode $\Psi_{-} $ does not depend upon value of $\eta $. 
\begin{center}
$\Psi_{-}= (\phi_{1}(x), i \phi_{2}(x), i \phi_{1}(x),\phi_{2}(x) )^{T}$, \vspace{1 mm}
\end{center}
 
where $\phi_{1}(x) =e^{-(x-\alpha-i\beta)^{2}/\sigma^{2}} + e^{-(x+\alpha-i\beta)^{2}/\sigma^{2}} + \eta  -(e^{-(x-\alpha+i\beta)^{2}/\sigma^{2}} + e^{-(x+\alpha+i\beta)^{2}/\sigma^{2}}) $; $\phi_{2}(x) =e^{-(x-\alpha+i\beta)^{2}/\sigma^{2}} + e^{-(x+\alpha+i\beta)^{2}/\sigma^{2}} + \eta  -(e^{-(x-\alpha-i\beta)^{2}/\sigma^{2}} + e^{-(x+\alpha-i\beta)^{2}/\sigma^{2}}) $. For $\Psi_{+}$, $J(x)$ \cite{foot} profile is constant and equals $\lambda_{so}$ for small values of $\beta$. Higher values of $\beta$ for given value of $\eta $ makes $J(x)$ un-physical and gives singularity for very large $\beta$ (See FIG. 1.). However, the solution is physically achievable for intermediate values of $\beta$ as the exchange energy i.e. $J(x)$ scale in G-type antiferromagnetic system, where spins align antiferromagnetically in the two sublattices of the buckled honeycomb lattice \cite{Long}, has order of $\sim 0.1$ eV.

As, it is apparent from the solution $\Psi_{+}$, that this can show the sub-Planck structure as it has all the interference terms required for sub-Planck. 
Wigner function for the above state can be written as: \vspace{1 mm} 

\begin{center}
    $W(x,p) =   \frac{N}{\pi h }\int_{-\infty }^{\infty}(\phi`^{*}(x+y)\phi`(x-y) + \eta \phi`^{*}(x+y) + \eta  \phi`(x-y) +\eta ^{2})  e^{\frac{2ipy}{h}} dy$ ,
\end{center}

\begin{figure}[h]
\includegraphics[width=8cm]{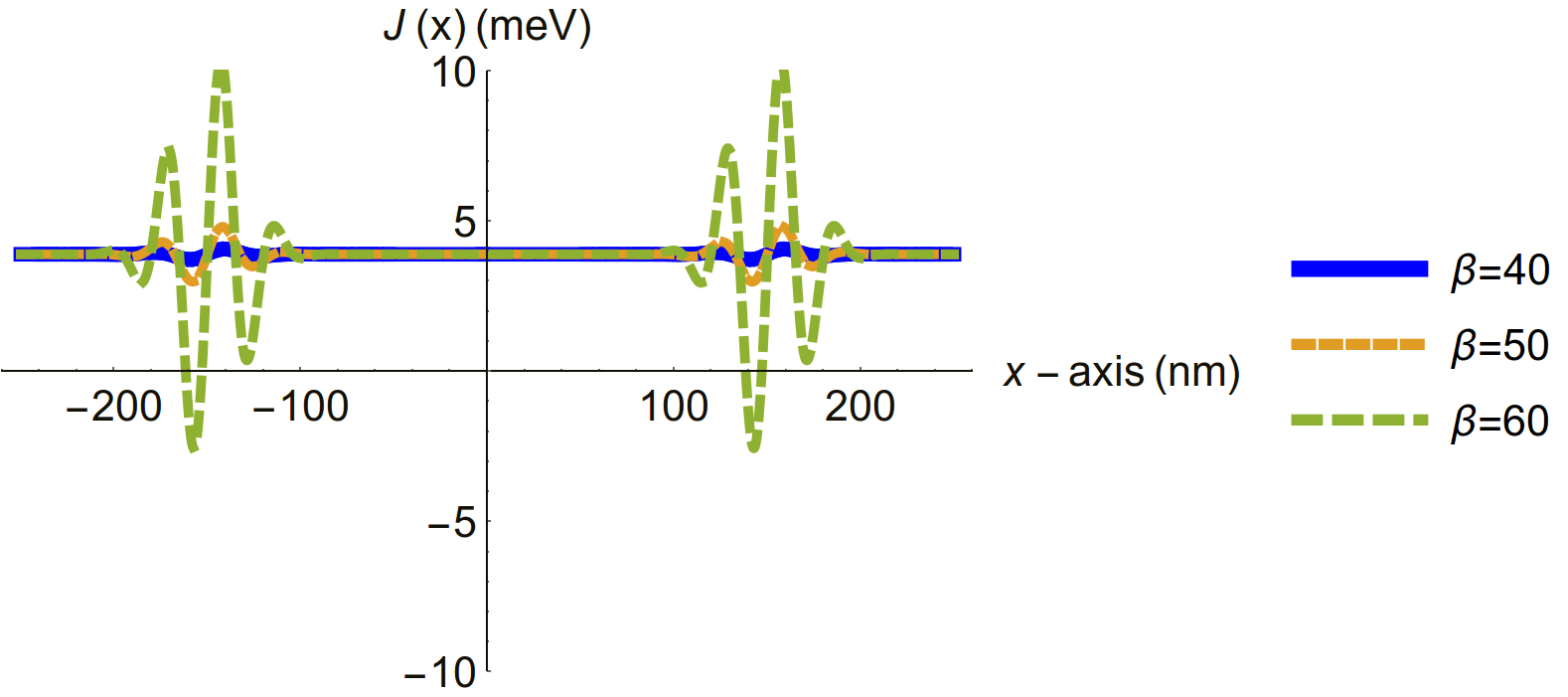} \vspace{2 mm}

\caption{Exchange energy between local magnetization and electron spin in silicene i.e., $J(x)$ profile variation along x-direction for  $\beta=40$, $\beta=50$, $\beta=60$. Value of $\alpha = 150$ nm.}

\end{figure}

where $\phi`(x) =  e^{-(x-\alpha-i\beta)^{2}/\sigma^{2}} + e^{-(x+\alpha-i\beta)^{2}/\sigma^{2}} +
e^{-(x-\alpha+i\beta)^{2}/\sigma^{2}} + e^{-(x+\alpha+i\beta)^{2}/\sigma^{2}}$. In the Wigner function shown in FIG. 2, we have not plotted the last term with $\eta ^{2}$, as it gives Dirac delta function. 
As mentioned in \cite{Zurek}, $ a\approx \frac{\hbar^{2}}{A}$ ( where $A=LP$ is phase space volume of the quantum state spread), determines the sensitivity of system to decoherence. FIG.4. clearly depicts the decrease in sensitivity as one gets closer to small momenta (y-axis).
On the other hand, spinor solution $\Psi_{-}$ shows symmetric sub-Planck structure. $J(x)$ for this solution always is independent of other parameters and has constant value $\lambda_{so}$. Wigner function in this case is independent of $\eta$ except for the $\eta^{2}$ term.
These solutions can simultaneously exist in the system by choosing the system parameters appropriately: small $\beta$ and large $\eta$. Also, we restore the TRS in the system as value of $J(x)$ is just $\lambda_{so}$, which cancel the diagonal term completely.
Although, it is possible to find other zero energy states, with different $J(x)$ profiles, appearance of sub-Planck structure only occurs for these solutions.

\begin{figure}[h]
\includegraphics[width=7cm]{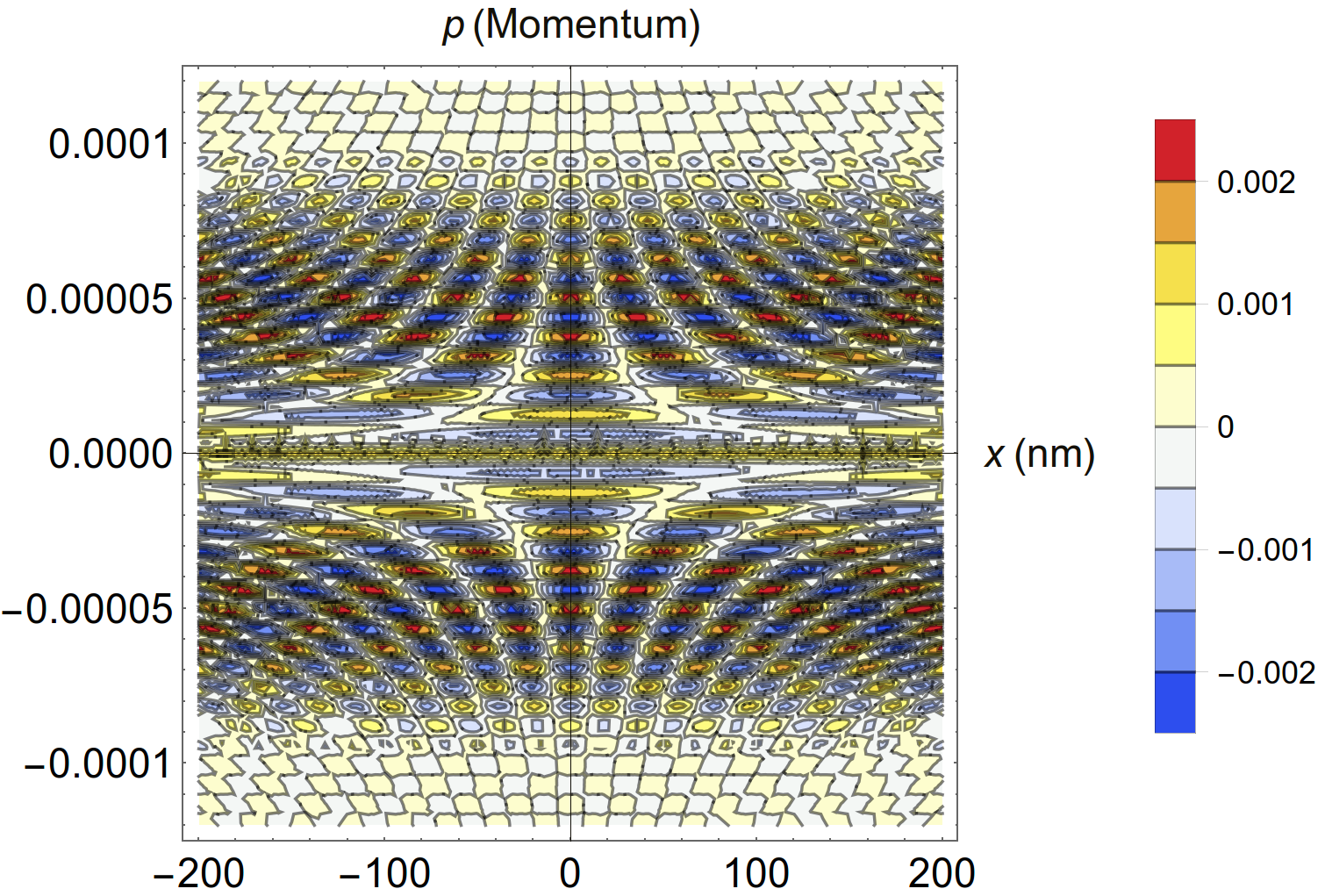}
\caption{Wigner function plot showing sub-Planck behavior in Silicene ground state $\Psi_{+}$ (un-normalized). $\alpha=250$ nm , $\beta = 30$ nm, $\sigma = 25$ nm $\lambda_{so} = 3.9 meV$, C = 10000. Color scale is adjusted to show all the points on contour. (In other contour plots which would follow, $\hbar$ has been taken of same order for plotting convenience).}

\end{figure}

\begin{figure}[h]
\includegraphics[width=7cm]{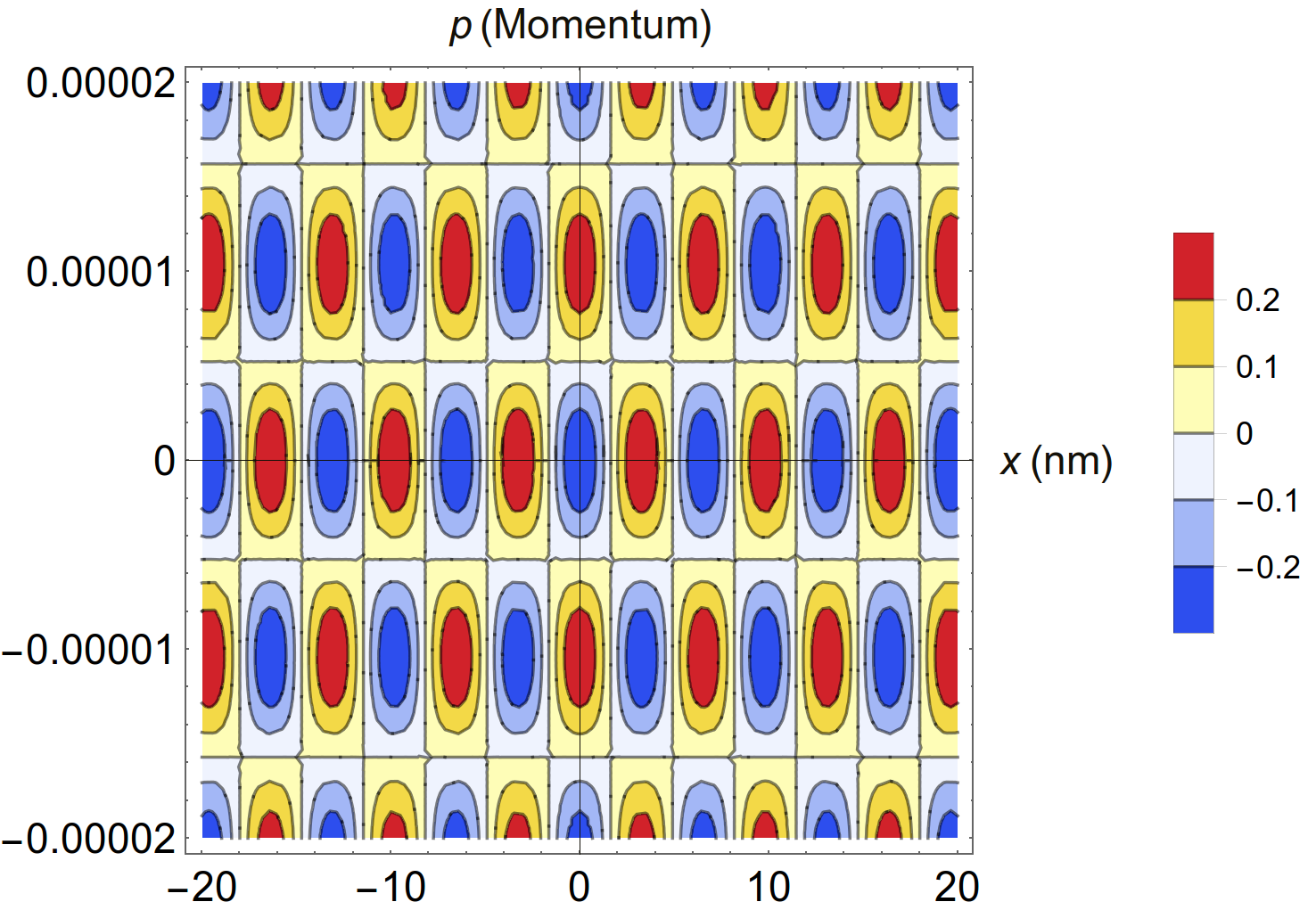}
\caption{Wigner function plot showing sub-Planck behavior in Silicene ground state $\Psi_{-}$, with parameters $\alpha=150$ nm , $\beta = 150$ nm, $\sigma = 25$ nm $\lambda_{so} = 3.9 meV$, $\hbar = 10^{-12}$. }

\end{figure}

\section{\label{sec:level1}inhomogenous gate electric field}
Now we identify and investigate another zero energy state in silicene under the application of suitably chosen out of the plane electric field. Appropriate tuning of electric field profiles allows superposition of Gaussian functions as the ground state of the system \cite{Asim}. Kitten like state have been shown in various quantum mechanical systems under different profiles \cite{Ghosh,PKP,Sayan}, exhibiting sub-Planck structure. Ezawa showed a Gaussian shifted in position can exist as zero energy mode \cite{ezaw} in silicene under application of linear electric field . As silicene modes are topological in nature only between electric field range $\pm E_{c}$, and ordinary insulator outside this range. Considering electric field of form :  

\begin{center}
   $E_z =\frac{1}{l\sigma^{2}}\frac{(\lambda_{so} \sigma^{2} + v_{f}(x+\alpha)h +(v_{f}h(e^{\frac{2x\alpha}{\sigma^{2}}})) )}{(e^{\frac{2x\alpha}{\sigma^{2}}} cos(\frac{(x-\alpha)\beta}{\sigma^{2}}) + cos(\frac{(x+\alpha)\beta}{\sigma^{2}}))} $\\ 
\end{center}

allows a zero energy state with sum of two shifted Gaussians centered around  $\pm\alpha$ (See FIG. 4).
We seek zero energy solution with particle hole symmetry.  Defining $\varphi(x) = e^{-\frac{(x-\alpha-i\beta)^{2}}{2\sigma^{2}}}  + e^{-\frac{(x+\alpha+i\beta)^{2}}{2\sigma^{2}}} + e^{-\frac{(x-\alpha+i\beta)^{2}}{2\sigma^{2}}} + e^{-\frac{(x+\alpha-i\beta)^{2}}{2\sigma^{2}}}$, the solution is $N (\varphi ,i\varphi ,0,0)^{T}$ (N is normalization constant). This state is maximally separable state in the four-dimensional subspace of total Hilbert space (spin, orbital, space). Pure spin current with a specific polarization can flow under this type of potential profile. FIG.4 and 5 shows the Electric field profile and sub-Planck structure for this state. Wigner function calculation for such relativistic type state has been done using approach mentioned in \cite{Campos}. This electric field profile can be implemented using schemes in transistor based devices \cite{Konde} . 

Interaction of such states with another qubit through $\sigma_{z}\otimes\sigma_{z}$ interaction, (which is common to interacting spin systems) undergoes decoherence free evolution. The four dimensional Hilbert sub-space is a decoherence free sub-space (DFS) \cite{Benatti}. However, considering more realistic  term $H`$ of the form $\lambda(\sigma_{z} \otimes \mathbb{I}_{2}) (-i\hbar\partial _{x})$, which induces displacement in phase space \cite{Zurek}
in the set up mentioned in section 5, can lead to decoherence.

\begin{figure}[h]
\includegraphics[width=7cm]{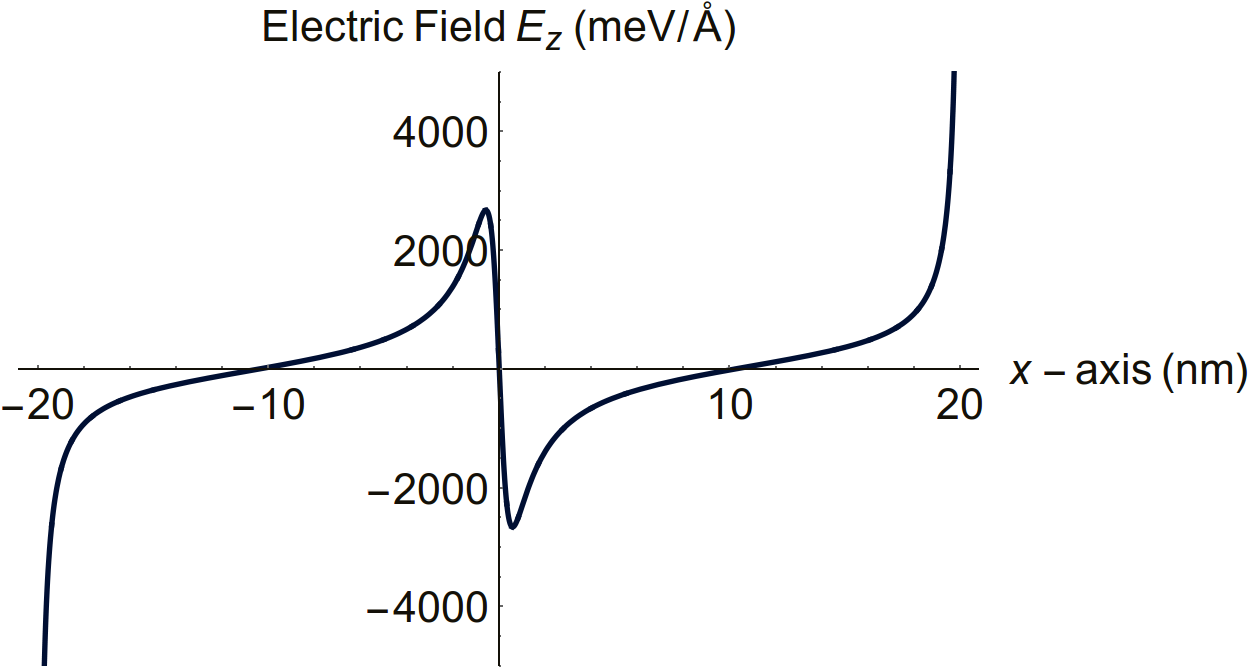}
\caption{Electric field profile along x-direction, for values $\alpha =10$ nm, $\beta = 10$ nm and $\sigma= 8$ nm. Electric field units are taken in $meV/\AA$  and x-axis in nm. Electric field has singularities on both sides of x-axis.}

\end{figure}

\begin{figure}[h]

\includegraphics[width=7cm]{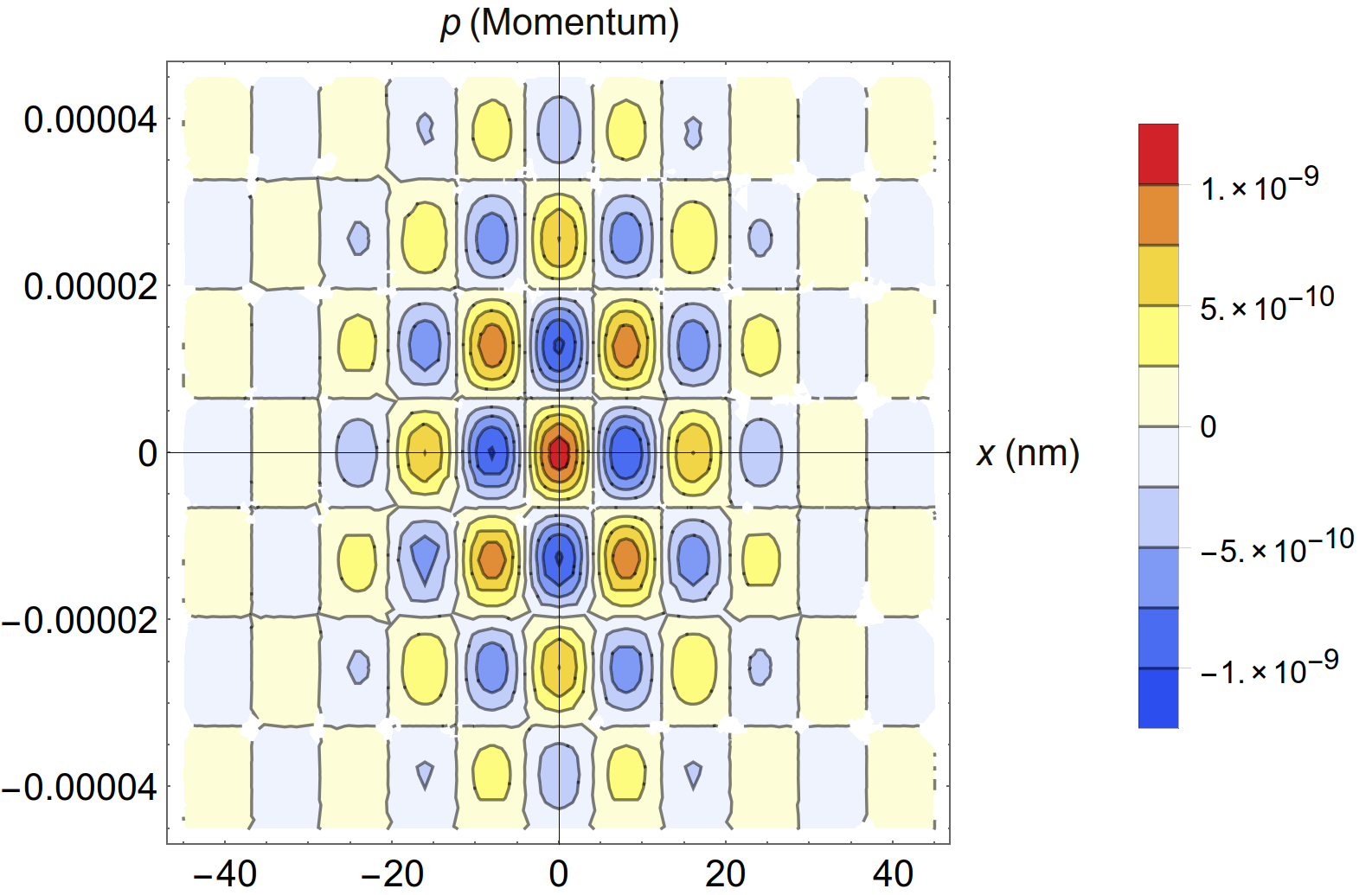}
\caption{Contour diagram showing sub-Planck structure in silicene under inhomogeneous electric field  (mentioned above). Values of $\alpha = 120$ nm,  $\beta= 120$ nm, $\sigma = 25$ nm, $\lambda_{so} = 3.9 meV$. }
\end{figure}

\section{n- qubit interaction and decoherence}

This section describes the interaction between silicene zero energy modes (mentioned above) and an n-qubit system. Silicene here has been considered as an environment. Total Hamiltonian is chosen to have this form:  

\begin{center}
   $ H = \sum_{i=1}^{n}c_{i}\sigma_{z}^{i} \otimes H^{`} +  \mathbb{I}_{2^{n}} \otimes H_{env}$ 
\end{center}

Where $\sigma_{z}^{i} = \mathbb{I}_{2}^{1}\otimes \mathbb{I}_{2}^{2} \otimes ....\otimes \sigma^{i}_{z}\otimes...\otimes\mathbb{I}_{2}^{n}$  ;
$H` = \lambda(\sigma_{z} \otimes \mathbb{I}_{2})$  
and $H_{env}$ is silicene Hamiltonian. This type of Hamiltonian is considered for dephasing only, as mentioned in \cite{Mac}.
Here, $c_{i}$'s are coupling coefficients for different qubits, which are non interacting. Also, initial state of the system (n-qubits) and environment (Silicene) is considered to be 

\begin{center}
    $\Ket{\Psi(0)} = \sum_{i} a_{i}\Ket{i}_{2^{n}} \otimes \Ket{\Psi_{-}}$ 

\end{center}

Where the first term represents an arbitrary superposition of eigenstates of $\sigma_{z}$ for n-qubits. At later instant $t$, the total state can be written as:  

\begin{center}
    $\Ket{\Psi(t)} = \sum_{i} a_{i}\Ket{i}_{2^{n}} \otimes \Ket{E_{i}}$
\end{center}

In order to calculate decoherence, we can write the reduced density matrix for the system at later instant of time $t$. The off diagonal components of a density matrix represents coherence. For n-qubit system there will be $^{2^n}\textrm{C}_{2}$ independent off-diagonal elements. These off-diagonal elements are overlap integrals of  different $\Ket{E_{i}}'s$ . For a single qubit system there is only one overlap integral, which takes the form:  \\

\begin{center}
    $I = \int_{-\infty }^{\infty}dx \sum_{s=1}^{4}\Bra{n_{s}} \Ket{E_{1}} \Bra{E_{2}}\Ket{n_{s}}$

\end{center}

Where $\Ket{E_{1}} = e^{-i\frac{t}{h}(H_{env} + c_{1}H^{`})} \Ket{\Psi_{-}}$ and $\Bra{E_{2}} = \Bra{\Psi_{-}} e^{i\frac{t}{h}(H_{env} -c_{1}H^{`})} $.  \\

Value of this overlap integral is a measure of decoherence. If this integral is just a phase then system undergoes coherent evolution. Also, if it takes value zero then system goes under complete decoherence. Compared to the state mentioned in section III, we get periodic information revival for $\sigma_{z}\otimes\sigma_{z}$ interaction. The value of the overlap integral $I$ for a single qubit system is found to be: 

\begin{center}
    $I = Cos(2\lambda c_{1}\frac{t}{h})$ 
\end{center}

Owing to perfect entanglement of spinor basis, decoherence is time-periodic for $\sigma_{z}\otimes\sigma_{z}$ type of interaction with environment being in initial state $\Ket{\Psi_{-}}$. Also, for n-qubits decoherence remains time periodic however minimum time period  available varies as $\frac{T}{n}$ (considering all $c_{i}$ to be same i.e. $c_{1}$), where $T$ is the time-period for single qubit case. Same behaviour follows for non zero $\eta$ term. Also, $\Psi_{+}$ yields the same result for this type of interaction.

\section{decoherence under Phase space shift}

In the above section interaction term only perturb spin degrees of freedom. Now we consider a term, whose effect is to displace the environment state in phase space as well. Let $H`$ to be $\lambda(\sigma_{z} \otimes \mathbb{I}_{2}) (-i\hbar\partial _{x})$ in the set up mentioned in last section. This type of interaction term may arise, if silicene is deposited on ferromagnet (110) plane, with zinc blende structure. As $H`$ commutes with $H_{env}$, the overlap integral for single qubit case can be written  in a straight forward manner for state $\Psi_{-}$ as: \\

\begin{center}
    $I = -4 \int_{-\infty }^{\infty}dx B(x+c_{1} \lambda t) B(x-c_{1} \lambda t)$ 
\end{center}

Where $B(x)$ is expression from section 4, with $C=0$ \\

\begin{figure}[h]
\includegraphics[width=7cm]{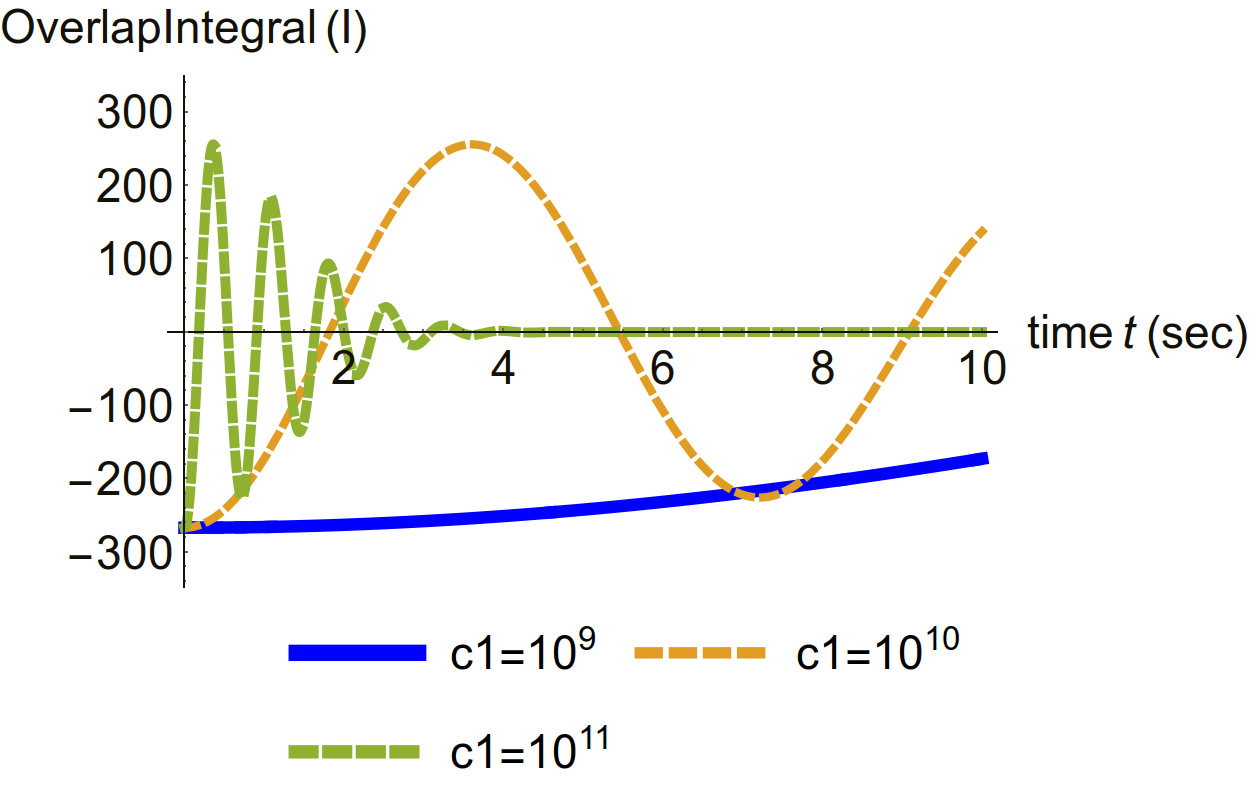}
\caption{Time dependence of the overlap integral I for different orders of coupling constant c1, for a single qubit interacting with $ \Psi{-}$ state. Overlap integral has been plotted in arbitrary units and time in seconds.} 
\end{figure}

\begin{figure}[h]
\includegraphics[width=7cm]{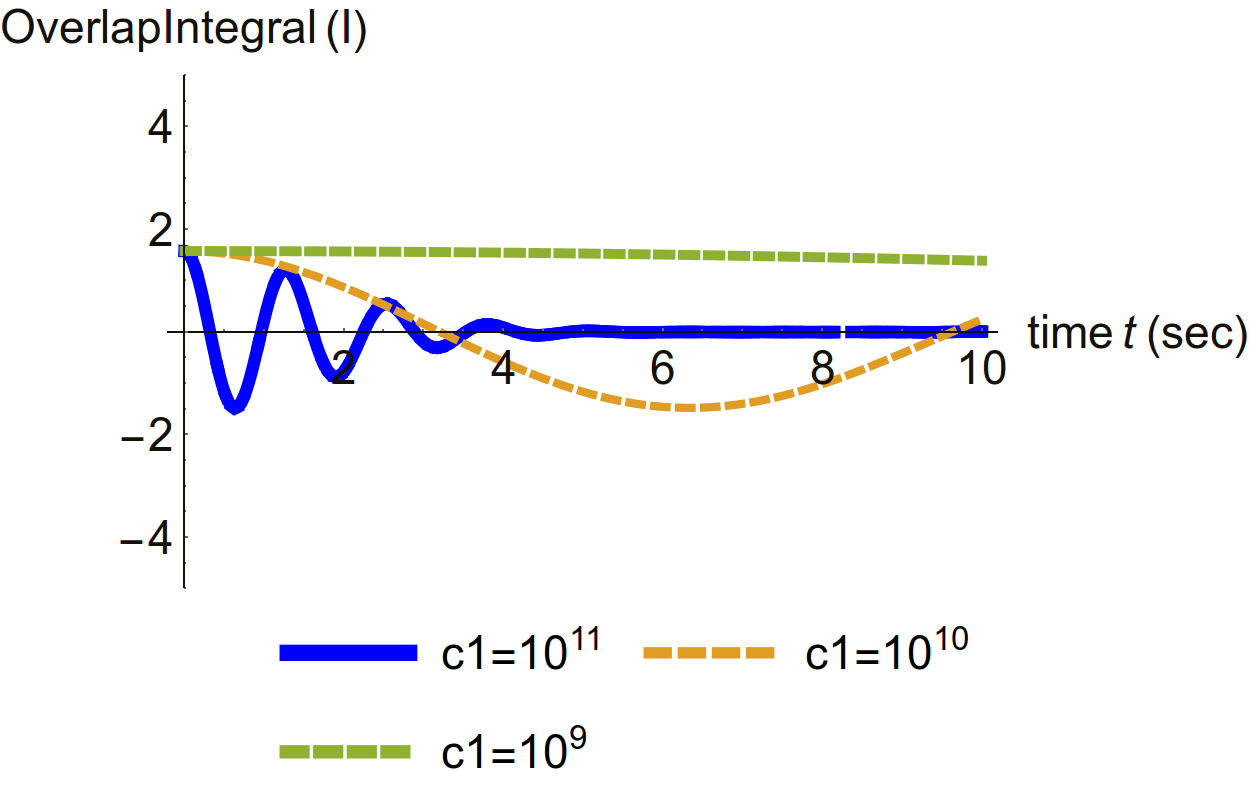}
\caption{Time dependence of the overlap integral for $\Psi_{+}$ state, for different coupling constants orders (same as above). This mode is more robust compared to the other, as the integral value decays slower than $\psi_{-}$.} 
\end{figure}

Majorana modes can also undergo decoherence \cite{Ng}, both in interacting and non-interacting scenario.  Here, we have non-interacting case with different coupling strengths. As clear from the FIG.6., for larger coupling coefficients $(10^{11})$ the integral vanishes quickly, and hence complete decoherence. For intermediate values  $(10^{10})$ it oscillates with decaying amplitude and therefore increasing decoherence. While for small coupling amplitudes  $(10^{9})$ system remains coherent. Hence, $\Psi_{-}$ would lead to decoherence as coupling strength increases. On the other hand, $\psi_{+}$ state is more robust, and remains decoherence free under time evolution for longer period (FIG. 7.), when interacted with a single qubit. It is straight forward to observe that the number of qubits n that can interact with the system is restricted owing to coupling coefficients and time of computation required. Still, such large values of coupling coefficients can allow large number of qubits to interact. For n-qubit case, one only needs to consider fastest decaying overlap integral. As shift in the Gaussians varies linearly with number of qubits. The fastest decaying integral can be written as below, which would replace $c_{1}$ in the integral with $c_{1}$$n^{2}$.: 
\begin{center}
    $I = \pm 4 \int_{-\infty }^{\infty}dx A/B(x+ n c_{1} \lambda t) A/B(x- n c_{1} \lambda t)$  
\end{center}
Here A(x) is a function defined for $\Psi_{+}$ and B(x) for $\Psi_{-}$ with $C=0$ in section IV.

\section{Conclusion}
In conclusion, we have explicitly demonstrated Majorana zero energy modes in silicene under the influence of opposite magnetic ordering. One of the modes is found to be a BIC and preserves it's coherence showing periodic revival, when interacts with a spin through $\sigma_{z}\otimes\sigma_{z}$ interaction. This may find applications in quantum computation, spintronics etc. Then under the influence of electric field zero energy mode is identified as an exact solution carrying pure spin current tunable with electric field. However, this mode rapidly lose it's coherence when interacted with external spin system. All these modes exhibit sub-Planck structure in the phase space. Both gate electric field and magnetic ordering set-ups are possibly achievable in laboratory conditions. Silicene can be used as an environment with controlled decoherence for an n-qubit system in different settings. Time evolution under phase shift interaction, leads to very slow decay of overlap integrals. Since $\Psi_{+}$ and $\Psi_{-}$ can exist simultaneously in the system for specific parameters range. It would be interesting to analyze the interacting case of Majorana modes. We also conclude that BIC state mentioned above has topological nature. Although, we have explicated silicene for these exotic features, however G-type AFM host material $ABO_{3}$ would be most suitable system to search for Majorana modes.

\section*{Acknowledgement} This work is supported by CSIR- JRF (Junior Research Fellowship), MHRD, Govt. of India. A.V.V. would like to thank Subhrajit Modak for his helpful discussions.

\end{document}